\documentclass[aps,prd,twocolumn,showpacs,amsmath,amssymb,nofootinbib,eqsecnum, preprintnumbers]{revtex4-1}

\usepackage{bbm}

\def\A{{\cal A}}

\def\obh{\Omega _{h}}
\def\ods{\Omega _{c}}

\def\J{{\cal J}}

\def\M{{\cal M}}

\newcommand{\n}[1]{\label{#1}}

\newcommand{\ba}{\begin{eqnarray}}
\newcommand{\ea}{\end{eqnarray}}

\def\ben{\begin{equation}}
\def\een{\end{equation}}
\def\bea{\begin{eqnarray}}
\def\eea{\end{eqnarray}}
\def\be{\begin{equation}}
\def\ee{\end{equation}}
\def\ft#1#2{{\textstyle{\frac{\scriptstyle #1}{\scriptstyle #2} } }}
\def\fft#1#2{{\frac{#1}{#2}}}
\def\cA{{\cal A}}

\newcommand{\nn}{\nonumber}

\begin{document}

\title{{Thermodynamic Volumes and Isoperimetric Inequalities for de Sitter Black Holes}}


\author{Brian P. Dolan}
\email{bdolan@thphys.nuim.ie}
\affiliation{Department of Mathematical Physics, National University of Ireland,
Maynooth, Ireland
\\
and\\
Dublin Institute for Advanced Studies, 10 Burlington Rd., Dublin, Ireland
}

\author{David Kastor}
\email{kastor@physics.umass.edu}
\affiliation{Department of Physics, University 
of Massachusetts,  Amherst, MA 01003}

\author{David Kubiz\v n\'ak}
\email{dkubiznak@perimeterinstitute.ca}
\affiliation{Perimeter Institute,
31 Caroline St. N. Waterloo Ontario, N2L 2Y5, Canada}

\author{Robert B. Mann}
\email{rbmann@uwaterloo.ca}
\affiliation{Department of Physics and Astronomy, University of Waterloo,
Waterloo, Ontario, Canada, N2L 3G1}
\affiliation{Perimeter Institute,
31 Caroline St. N. Waterloo Ontario, N2L 2Y5, Canada}

\author{Jennie Traschen}
\email{traschen@physics.umass.edu}
\affiliation{{\,\!}Department of Physics, University 
of Massachusetts,  Amherst, MA 01003}

\date{May 1, 2013}

\begin{abstract}
We consider the thermodynamics of rotating and charged asymptotically de Sitter black holes.  
Using Hamiltonian perturbation theory techniques, we derive three different
first law relations including variations in the  cosmological constant, and associated Smarr formulas that are satisfied by such spacetimes.
Each first law introduces a different thermodynamic volume conjugate to the cosmological constant.  We examine the relation between these thermodynamic volumes and associated geometric volumes in a number of examples, including Kerr-dS black holes 
in all dimensions and Kerr-Newman-dS black holes in $D=4$.  {We also show that the Chong-Cvetic-Lu-Pope solution of $D=5$ minimal supergravity, analytically continued to positive cosmological constant, describes black hole solutions of the Einstein-Chern-Simons theory and include such charged asymptotically de Sitter black holes in our analysis.}  
In all these examples we find
that the particular thermodynamic volume associated with the region between the black hole and cosmological horizons is equal to the naive geometric volume. 
Isoperimetric inequalities, which hold in the examples considered, are formulated for the different thermodynamic volumes and conjectured to remain valid for all asymptotically de Sitter black holes. In particular, in all examples considered, we find that for fixed volume of the observable universe, the entropy is increased by adding black holes. We conjecture that this is true
in general.
\end{abstract}

\preprint{DIAS-STP-13-1, pi-stronggrv-314}

\maketitle

\section{Introduction}

That black holes have a temperature proportional to their surface gravity and obey a version of the first law of thermodynamics is a notion that has been of continued interest for over three decades.   Studies of black holes that are asymptotically flat or anti de Sitter  (AdS) have been carried out in considerable detail, with many corroborative results 
indicating that such objects indeed behave as thermodynamic systems.   Thermodynamic equilibrium is straightforward to define, and (depending on the kind of black hole and its asymptotics) a variety of interesting phenomena emerge, such as large/small  AdS black hole phase transitions \cite{HawkingPage:1983}.  An interesting new development in the study of black hole thermodynamics was  the proposal that the  mass of an AdS black hole should be interpreted as the enthalpy of the spacetime.  This notion emerges from geometric derivations of the Smarr formula for AdS black holes 
which suggest that the cosmological constant should be considered as a thermodynamic variable analogous to pressure in the first law
\cite{CaldarelliEtal:2000, KastorEtal:2009, Dolan:2010, Dolan:2011a, Dolan:2011b, Dolan:2012, CveticEtal:2010, LarranagaCardenas:2012, LarranagaMojica:2012, 
Gibbons:2012, KubiznakMann:2012, GunasekaranEtal:2012, BelhajEtal:2012,  LuEtal:2012, SmailagicSpallucci:2012, HendiVahinidia:2012}. Further, this notion led to a $reverse$
isoperimetric inequality conjecture  \cite{CveticEtal:2010} for AdS black holes, which
says that for fixed thermodynamic volume, the entropy of an AdS black hole is maximized
for Schwarzchild AdS. Since black holes in a spacetime with positive $\Lambda$
are of interest in cosmology,
in this paper we study these issues for de Sitter black holes.

Studies of asymptotically de Sitter (dS) black holes are somewhat more sparse,
and need to confront two interesting and related issues. {It is fundamental to the first
law for asymptotically flat black holes that} there is a Killing field which is timelike
everywhere outside the black hole and hence can be used to define the mass. 
Adding a multiple of the rotation Killing field to this timelike Killing field gives
the generator of the horizon, and the Killing generator 
 defines the temperature. The same is true with negative cosmological constant.
However in de Sitter there is no Killing field that is timelike {everywhere outside the black hole, }
including infinity.  Although generalizations of the first law {in de Sitter} were  considered quite some time ago \cite{GibbonsHawking:1977}, {a further}  key issue is that an observer in any static patch of the space-time is between two horizons, each with its own distinct constant surface gravity.  It is consequently unclear how to understand the thermodynamics of {such} spacetimes, since there is neither thermodynamic equilibrium nor an asymptotic region where a conserved mass-energy can be defined\footnote{Some consideration has been given to defining a mass-energy-like quantity at future/past infinity in asymptotically dS spacetimes \cite{Balasubramanian:2001nb,Ghezelbash:2001vs,Ghezelbash:2002vz}. Positivity of a conformal mass {defined in this context} was shown in \cite{Kastor:2002fu}}.
Various approaches have been taken toward addressing this problem, yielding mixed results.

In this paper we consider the thermodynamics of  de Sitter black holes, focusing on the cosmological constant as a thermodynamic variable together with its conjugate potential.  To highlight the correspondence with classical thermodynamics we also work with the pressure $P$ related to the cosmological constant $\Lambda$ according to 
\be\label{P}
P=-\frac{\Lambda}{8\pi}\,,
\ee
as it would be in a perfect fluid stress-energy.
The potential conjugate to $P$ is a volume $V$, which we will call the thermodynamic volume as in \cite{Dolan:2010, CveticEtal:2010}.  

It was found in \cite{KastorEtal:2009} that the thermodynamic  volume of  a Schwarzschild-AdS black hole is equal to the geometric volume $V^\prime$ of the black hole interior,   computed using the full $D$-dimensional volume element on a $t$ equals constant slice.  The relation between thermodynamic and geometric  volumes was studied  in \cite{CveticEtal:2010} for a variety of charged and rotating AdS black holes.  Equality between the two volumes was found to hold only in the static case (including charge), while with rotation the volumes
 differ by a simple correction term.  It  was also conjectured in \cite{CveticEtal:2010} and verified for a variety of black hole spacetimes that the thermodynamic volume satisfies an inequality with respect to the black hole horizon area that is precisely the reverse of the isoperimetric inequality of Euclidean space.

In this paper we will carry out a similar exploratory study of thermodynamic and geometric volumes for asymptotically de Sitter  black holes.  An important new feature in this case is the need to distinguish between several different thermodynamic volumes.  The geometric derivation of the first law and Smarr formula for AdS black holes \cite{KastorEtal:2009} takes place on a timelike hypersurface stretching outward from the black hole horizon to spatial infinity.  For dS black holes, we will see that additional relations are obtained by alternatively considering  the region stretching outwards to infinity from the  black hole and cosmological horizons, or the region between the black hole and cosmological horizons.   Each of these introduces a different thermodynamic volume.
We denote the thermodynamic volumes associated in this way with the black hole and cosmological horizons by $V_h$ and $V_c$, while reserving $V$ for that associated with the region between the two horizons\footnote{Similar results have been obtained by Cai \cite{Cai:2002, Cai:2002b} and by Sekiwa  \cite{Sekiwa:2006} in more limited contexts (see also \cite{UranoEtal:2009} and \cite{Gibbons:2005vp}). A study of the thermodynamics of asymptotically flat and AdS rotating black holes that included both inner and outer horizons, which has elements in common with our approach,  was recently carried out \cite{ChenEtal:2012}.}.    The different thermodynamic volumes may be thought of as arising from different thermodynamic ensembles, {\it e.g.}, $V_h$ arises from varying the cosmological constant with the black hole horizon area held fixed, while $V_c$ is relevant if instead the cosmological horizon area is fixed.

Turning to specific examples, for Kerr-de Sitter black holes we find that $V_h$ and $V_c$ are 
equal to their geometric counterparts $V_h^\prime$ and $V_c^\prime$
only for  static spacetimes, each differing by a simple expression when the black hole
has angular momentum. These relations are the same as
in AdS \cite{CveticEtal:2010}.  However, for the region between the two horizons the
rotation-dependent differences cancel, so that $V=V^\prime$ holds in the rotating case as well.  In line with this result, we find that the thermodynamic volumes $V_h$ and $V_c$ satisfy reverse isoperimetric inequalities, while $V$ satisfies a true isoperimetric inequality.  These inequalities may be interpreted as bounds on combinations of the black hole and cosmological entropies at a fixed thermodynamic volume. In particular, in all examples considered, we find that for fixed volume of the observable universe, the entropy is increased by adding black holes. We conjecture that this is true in general.

Our results for charged rotating black holes are less general.  In four dimensions, one has the Kerr-Newman-de Sitter spacetimes and in this case we find results that simply extend those of the zero charge case.  In five dimensions we consider charged, rotating black holes {of $D=5$ Einstein-Chern-Simons theory with positive cosmological constant.\footnote{%
For a discussion of higher-dimensional charged rotating (A)dS black holes constructed numerically see, e.g., \cite{BrihayeDelsate:2008br}. 
}
We find that an exact solution for such black holes can be obtained} from the gauged supergravity solution of 
Chong, Cvetic, Lu, and Pope \cite{ChongEtal:2005b}, by analytically continuing the gauge coupling 
to obtain positive values of the cosmological constant.
We compute the thermodynamic volumes $V_h$, $V_c$ and $V$ for these spacetimes and verify that $V$ coincides with the geometric volume.  While we are able to analytically verify the isoperimetric inequality for a singly-rotating Chong-Cvetic-Lu-Pope black hole, because of the complexity of the expressions we have not been able to analytically complete the study of isoperimetric inequalities for the most general case. We have, however, strong indications from numerical
analysis that such inequalities remain satisfied.

The outline of our paper is as follows.
In the next section the three first laws and the corresponding Smarr formulae are derived using the Hamiltonian perturbation techniques. The thermodynamics of the
Kerr-deSitter black holes in all dimensions is discussed in Sec.~\ref{kerrds} while the isoperimetric inequalities involving the corresponding thermodynamic volumes are discussed in Sec.~\ref{isop}. The charged de Sitter black holes, their thermodynamics and isoperimetric inequalities are studied in Sec.~\ref{charged}. 
Sec.~\ref{compr} displays results on compressibility and speed of sound for black hole horizons.
Sec.~\ref{sum} is devoted to the summary. Appendix~\ref{Nari} discusses the Nariai limit and the corresponding thermodynamic volume in between the horizons.

\section{First Law and Smarr Formula}\label{firstlaw}

In this section we apply the Hamiltonian formalism of general relativity  \cite{Traschen:1985,SudarskyWald:1992,TraschenFox:2004} to derive the first law 
relations for
rotating dS black holes  including variations in the cosmological constant.  As noted above, we will actually find three different first law relations associated with different choices for the hypersurface
and the Killing vector
 which enter the derivation.  Two of these relations are linearly independent.  Each version of the first law leads to an associated Smarr relation via an overall scaling.

\subsection{Gauss' law for perturbations with $\delta \Lambda$}

The essence of this Hamiltonian perturbation theory method is as follows.  In Einstein  gravity with cosmological constant, suppose one has a black hole solution with a Killing field.  Now consider solutions that are perturbatively close to this background solution, but are not required to have the original Killing symmetry.  The linearized Einstein constraint equations on a hypersurface can be expressed in the form of a Gauss' law (see \cite{Traschen:1985}), relating a boundary integral at infinity to a boundary integral at the horizon.
The physical meaning of this Gauss' law relation depends on the choice of Killing field, as well as on the choice of hypersurface.   Taking the generator $l^a$ of a Killing horizon, together with an appropriate choice of a spacelike hypersurface, yields the usual first law for   variation of the mass \cite{SudarskyWald:1992} for  asymptotically flat or AdS black hole spacetimes.

Assume we have a foliation of a spacetime by a family of hypersurfaces 
denoted by $\Sigma$
 and the unit timelike normal to the hypersurfaces $n^a$, $n\cdot n=-1$. The spacetime metric can then be written as
\begin{equation}\label{metricsplit}
g_{ab} = s_{ab}-n_a n_b\,,
\end{equation}
where $s_{ab}$ is the metric on the hypersurfaces $\Sigma$ and  satisfies $s_a{}^b n_b =0$.
As usual, the dynamical variables in the Hamiltonian formalism  are
the  metric $s_{ab}$ and its canonically conjugate momentum $\pi ^{ab} =-\sqrt{s} (K^{ab}-Ks ^{ab} )$.  Here  {$K_{ab} =s_a{}^c \nabla _c n_b$}
is the extrinsic curvature of a hypersurface ${\Sigma}$ and $K$ stands for its trace, $K=K^a{}_a$. (Similarly later we define $\pi=\pi^a{}_a$ and $h=h^a{}_a$.)  We consider Hamiltonian evolution
along the vector field $\xi ^a$, which can be decomposed into its components normal and tangential to $ {\Sigma}$ according to
\begin{equation}\label{xidecomp}
\xi^a=Fn^a +\beta^a\,,
\end{equation}
with $F=-\xi\cdot n$ denoting the lapse function and $\beta^a$ the shift.
The gravitational Hamiltonian 
which evolves
the system along $\xi^a$ is then given by ${\cal H} =FH+\beta^a H_a$ with
\ba
H&\equiv& -2G_{ab} n^a n^b =  -\, R^{(D-1)}  + {1 \over |s|}  \Bigl({\pi ^2 \over D-2 } - \pi^{ab} \pi_{ab} \Bigr)\,,\label{hamconstraint}\nonumber\\
H_b&\equiv& -2G_{ac} n^a s^c _b=-2\,  D_a (|s|^{-{1 \over 2}} \pi^{ab} )\,.\label{momconstraint}
\ea
Here $R^{(D-1)}$ is the scalar curvature for the metric $s_{ab}$ and $D_a$ is the derivative operator on the hypersurface $ {\Sigma}$.
With a cosmological constant stress-energy ${8\pi} T^a _b =-\Lambda g^a _b $, the constraint
equations become
\begin{equation}
H =- {2} \Lambda\,,  \quad H_b =0\,.
\end{equation}

Let $g_{ab}$ be a  solution to the Einstein equation with cosmological constant, and assume that 
$\xi^a$ is a Killing vector of $g_{ab}$.
Now let the metric $\tilde g_{ab} =g_{ab} +\delta g_{ab}$ be the linear approximation to 
another solution to the  Einstein equations with cosmological constant 
$\Lambda +\delta\Lambda$. 
Denote the Hamiltonian data for the background metric by $s_{ab} ,\pi^{ab}$, the corresponding perturbations to the data by $h_{ab}=\delta s_{ab}$ and $p^{ab}=\delta \pi ^{ab} $, and the linearized Hamiltonian and momentum constraints by $\delta H$ and $\delta H_a\,.$   

It was shown in \cite{Traschen:1985,SudarskyWald:1992,TraschenFox:2004} that
 a particular linear combination of the perturbed constraints can be written as a total derivative,
$F\delta H + \beta^a\delta H_a= D_a B^a$,
where the vector $B^a$ is given by
\ba\label{gaussvector}
 B^a [\xi ] &=&   F(D^a h - D_b h^{ab})  - h D^a F + h^{ab} D_b F \nonumber\\
&&+ {1 \over \sqrt{|s|} } \beta^b\bigl(\pi^{cd} h_{cd} s^a{}_b - 2 \pi^{ac} h_{bc} -2p^a{}_b \bigr)\,.\qquad
\ea
On the other hand, since $h_{ab}, p^{ab}$ solve the linearized  constraint equations with the cosmological constant perturbed by $\delta \Lambda$ 
we also have
$F\delta H + \beta^a\delta H_a= {2} \delta \Lambda \xi^a n_a =- {2} \delta \Lambda F$
and therefore
\begin{equation}\label{gauss}
D_a B^a =  - {2} F\delta \Lambda\,.
\end{equation}  
This has the form of a Gauss' law relation with source proportional to $F\delta \Lambda$.
 In \cite{Kastor:2008xb,KastorEtal:2009} it was shown that since  $\xi^a$ is a Killing vector, this source
 may also be written as a total derivative. We define the Killing potential $\omega^{ab}$
 associated with $\xi ^a$ to be an antisymmetric tensor satisfying
 \begin{equation}\label{omegadef}
 \nabla _c \omega ^{cb} = \xi ^b\,.
 \end{equation}
 The lapse function
 can now be written as
$F=-D_c (\omega ^{cb} n_b )$
and the relation (\ref{gauss}) becomes 
 \be\label{altgauss}
 D_a ( B^a   -{2}\delta \Lambda \omega ^{ab} n_b ) =0\,.
 \ee
 
Let $\hat V$ be a volume contained in $\Sigma$ which has  inner and outer boundaries
$\partial \hat V_{in, out}$. Integrating the differential relation (\ref{altgauss}) over $\hat V$ gives
\begin{equation}\label{gaussint}
\begin{gathered}
\int_{\partial \hat V_{out}} da r_c \left( B^c [\xi ]  - {2}\delta \Lambda \omega ^{cb} n_b \right) \qquad\qquad\\
\qquad \qquad \qquad = \int_{\partial \hat V_{in}} da r_c \left( B^c [\xi ]   - {2} \delta \Lambda \omega ^{cb} n_b \right)\,.
\end{gathered}
\end{equation}
Here we have let $r_c$ denote the unit  normal on each boundary, with
the convention that $r^c$ points into $\hat V$ on the inner boundary and out of
$\hat V$ on the outer boundary, {\it i.e.} ``towards infinity'' on both boundaries.
In the following we will consider different cases,  in which these boundaries may be taken to be at the black hole horizon, the de Sitter
horizon, and at infinity.

The different boundary integrals have important geometrical meanings.
 For an asymptotically flat or AdS black hole the variation in the ADM mass and
 angular momentum are given by the boundary integrals at infinity using the time translation
$ (\partial/\partial t)^a$
 and rotation $ (\partial/\partial \varphi)^a$ Killing vectors, respectively. Also, the integral of the boundary term
 over the horizon using  the horizon generating Killing vector  is proportional to the surface gravity times the change in area of
 the black hole. Explicitly,
 \begin{eqnarray}
 16\pi \delta \M  &=&
 -\int_\infty da r_c B^c  [{\partial/\partial t}]\,,  \label{flatdm} \\  \cr
  16\pi \delta \J  &=&  \int_\infty da r_c  B^c  [{\partial/\partial \varphi} ]\,, \label{flatdj} \\ \cr
   2\kappa_{h} \delta \A_{h}  &=&  -\int_{h} da r_c  B^c  [ {\partial/\partial t+ \obh \partial/\partial \varphi} ]\,,  \label{bhhor}
 \end{eqnarray} 
 where $\M$ and $\J$ are the ADM mass and momentum, $\kappa _{h}$
 is the surface gravity, and $\A_{h} $ is the area of the black hole. 
 When evaluating the boundary term on the black hole horizon we have assumed that it is a bifurcate
 Killing horizon and made use of the fact that the Killing generator
 vanishes on the bifurcation sphere.
 
Hence for $\Lambda \leq 0$, we see that taking $\hat V_{in}$ to be the black hole horizon and
$\hat V_{out} $ the boundary at infinity,  
equation (\ref{gaussint}) with ${\delta}\Lambda \equiv 0$ gives 
the usual first law $\delta \M =\kappa_{h}\delta \A_{h}/(8\pi) +\obh \delta \J$. 
We now turn to evaluating the boundary terms in (\ref{gaussint})
for different choices of $\hat V_{in}$ and $\hat V_{out}$ and now with the variation in the cosmological constant assumed to be non-vanishing.

\subsection{First law in de Sitter with $\delta \Lambda$ and Smarr formula}\label{sec:Smarr}
There are several features of dS spacetime that are different from  AdS or Minkowski that
make the nature of a first law in dS distinct. Infinity in an asymptotically dS spacetime
is a spacelike surface. Cosmological space-like slices asymptote to one corner of the conformal
diagram. In Schwarzchild-like coordinates, surfaces of constant $t_{schw}$ become
timelike outside the cosmological horizon, and end at spacelike infinity. Likewise,
the static Killing field $\xi ^a = (\partial / \partial t_{schw} )$ is space-like outside the horizon
to infinity. One can use this Killing field to define an ADM charge at infinity, but
it does not have the usual interpretation of a mass in the sense of Noether's theorem.
On the other hand $\xi^a$ is timelike between the black hole and cosmological horizon.
This situation leads to three natural first law constructions in a black hole 
dS spacetime---from the black hole horizon to infinity, from the cosmological horizon to infinity,
and between the two horizons. Only two of these are independent, as the construction
between the horizons is the difference between the other two.

We first consider the first law construction outlined above for the region $\hat V$ contained between the black hole and cosmological horizons, rather than running the integrals out to infinity as in the asymptotically flat or AdS cases.
This  will give a relation between the variations of the two horizon areas, and does
not include a mass parameter.

Let us first consider the contribution generated by the 
Killing vector ${(\partial/\partial\varphi)^a}$.
The variation ${\delta J}$ in the  angular momentum 
is  still given by equation (\ref{flatdj}), but now 
evaluated in an asymptotically
de Sitter spacetime.  Using equation (\ref{gaussint}) with the inner boundary taken
to be the de Sitter horizon and the outer boundary at infinity one has
\begin{equation}\label{jchange}
  16\pi \delta \J  =  \int_\infty da r_c  B^c  [{\partial/\partial \varphi} ] =\int_{dS} da r_c  B^c  [{\partial/\partial \varphi} ]\,. 
  \end{equation}
 So $\delta J$
is also given by evaluating the boundary integral 
on the de Sitter horizon. 
The Killing potential term does not {contribute because} we can connect the boundaries with
a slice {having} normal $n_a $ proportional to $ \nabla _a t$, and hence {the lapse function}
 $F= -n_a (\partial /\partial \varphi) ^a$ in equation (\ref{gauss}) vanishes.

Substituting the generator of the  black hole horizon ${(\partial/\partial t)^a+ \obh (\partial/\partial \varphi)^a}$  into the boundary 
integrand on the black hole gives the same result as in
 the asymptotically flat case, so that equation (\ref{bhhor}) continues
to hold. The generator of the  de Sitter horizon $ {(\partial/\partial t)^a+ \ods (\partial/\partial \varphi)^a}$  substituted into the boundary
term on the de Sitter horizon  gives a similar result
\begin{equation}\label{dshor}
   2|\kappa _{c}| \delta \A_{c} =  \int_{dS} da r_a  B^a  [ {\partial/\partial t+ \ods \partial/\partial \varphi} ]\,,  
\end{equation}
{with appropriate care taken for the signs}. We have introduced the explicit 
absolute value signs on $\kappa_c$ for clarity, since the definition of surface 
gravity gives $\kappa _c <0$.
Finally, noting that the boundary vector (\ref{gaussvector})  is linear in the Killing field, {one finds that the integral of the boundary term for the black hole horizon generator over the de Sitter horizon may be written as}
\bea\label{dshortwo}
 \int_{dS} da r_a  &&B^a  [ {\partial/\partial t+ \obh \partial/\partial \varphi} ]  = \\ \nonumber &&    2|\kappa _{c}| \delta \A_{c} + 
16\pi (\obh -\ods )\delta \J\,.
\eea

The derivation of the first law is then completed by substituting the generator of the black 
hole horizon ${\partial/\partial t+ \obh \partial/\partial \varphi} $ into the Gauss' law integral
identity (\ref{gaussint}) and using (\ref{bhhor}) and (\ref{dshortwo}) {to arrive at}
\begin{equation}\label{first}
 { \kappa_{h}\delta \A_{h} \over 8\pi}+   {|\kappa_{c}|\delta \A_{c} \over 8 \pi } 
+ \frac{V \delta \Lambda}{8\pi}   +(\obh - \ods )\delta\J   =0\,,
\end{equation}
where the positive [cf. expression \eqref{VSchw} below] thermodynamic volume  $V$
is defined by 
\begin{equation}\label{vdef}
V=- {\left(  \int _{dS} da r_c n_b \omega ^{cb} -\int _{bh} da r_c n_b \omega ^{cb} \right)}\,.
\end{equation}
%

The Smarr relation for rotating black holes in de Sitter spacetime follows by integrating the first law
under a scaling transformation. The scaling dimensions of $\A$ and $\J$ are $D-2\,,$ while
$\Lambda$ has dimension $-2$, giving
\begin{equation}\label{smarr}
 {\kappa_{h} \A_{h} \over 8\pi}+   {|\kappa_{c}|\A_{c} \over 8 \pi }  
+(\obh - \ods )\J  =  \frac{V\Lambda}{4\pi (D-2)}\,.
\end{equation}
%

Let us now identify the positive cosmological constant $\Lambda$ with the negative thermodynamic pressure as in \eqref{P}, $P=-\frac{\Lambda}{8\pi}<0$,
commensurate with the definition in the anti de Sitter case \cite{GunasekaranEtal:2012}.
With this definition we get the following  forms for the first law and the Smarr relation 
respectively (including the possibility of more than one rotation parameter) 
\ba
0&=&T_{h} \delta S_{h}+T_{c}\delta S_{c}\nonumber\\
&&\qquad  \quad +\sum_i (\Omega^i_{h}-\Omega^i_{c})\delta J^i-V \delta P\,,\label{first2}\\
0&=&T_{h}S_{h}+ T_{c}S_{c}\nonumber\\
&&\qquad \quad + \sum_i (\Omega^i_{h}-\Omega^i_{c})J^i+\frac{2}{D-2}PV\,,\label{smarr2}
\ea
where we identified the horizon area with the entropy $S=\frac{\A}{4}$ and 
have defined the cosmological and black hole temperatures to be the positive quantities
 $T_{c}=\frac{|\kappa_{c}|}{2\pi}$ and  $T_{h}=\frac{\kappa_{h}}{2\pi}$.

\subsection{Going to infinity}\label{sec:2horizons}
One can essentially repeat the derivation of section~\ref{sec:Smarr} in two additional cases: i) one takes the slice to go from the  black hole horizon  to infinity 
(passing through the dS horizon), and ii) the slice extends from the cosmological horizon to
 infinity. Consequently one obtains two additional first law relations and   associated Smarr formulae, one for each horizon. In both cases, the expressions involve a quantity $M$ which  would be the ADM mass in the flat and AdS cases. In the dS case however, such a quantity is ``conserved in space'' (rather than in time) due to the spacelike character of the Killing field $ (\partial/\partial t)^a$ in the region near infinity.  Keeping this (important)
  distinction in mind, we shall refer to $M$ as the ``mass'' \cite{Ghezelbash:2001vs},
  and use it as a book-keeping device.

For the black hole horizon we {then get the first law relation}
\be
\delta M=T_{h} \delta S_{h}+\sum_i (\Omega^i_{h}-\Omega^i_{\infty})\delta J^i\, {+\, V_{h}\delta P\,,}\label{firstBH}
\ee
where the quantities $\Omega_\infty^i$ allow for the possibility of a rotating frame at infinity (see, {\it e.g.}, reference \cite{Gibbons:2005b}), 
$\delta M$ is given via the boundary integral
\be
16\pi\, \delta M  =- 
 \int_\infty da r_c \left( B^c  [\partial/\partial t]\, {+2\,\delta\Lambda\,  \omega^{cd}_{dS}\, n_d } \right )\,, \label{Mchange}
\ee
and the thermodynamic volume $V_h$ is given by the expression
\be\label{vdefBH} 
V_{h}=\int _{\infty} da r_c n_d( \omega ^{cd}{-\omega_{dS}^{cd})} -\int _{bh} da r_c n_d \omega ^{cd} \, .
\ee
The quantity $\omega_{dS}^{ab}$ appearing in equations (\ref{Mchange}) and (\ref{vdefBH})
is the Killing potential of the background de Sitter spacetime, without the black hole.  These contributions serve to make each of the quantities 
$\delta M$ and $V_h$ finite.  See reference \cite{KastorEtal:2009} for an in depth discussion of this point in the asymptotically AdS case.  The corresponding Smarr formula, obtained from the first law by overall scaling, is then given by
\be
\frac{D-3}{D-2}M=T_{h}S_{h}+\sum_i (\Omega^i_{h}-\Omega^i_{\infty})J^i{-\frac{2}{D-2}PV_{h}\,,}\label{smarrBH}
\ee
We see from the first law \eqref{firstBH} that the thermodynamic volume $V_h$ may be interpreted as the change in the mass under variations in the cosmological constant with the black hole horizon area and angular momentum held fixed.

For the cosmological horizon, {we use the Gauss' law relation (\ref{gaussint}) with the corresponding horizon generating Killing vector 
$(\partial/\partial t)^a +\Omega_c(\partial/\partial \varphi)^a$ to obtain the first law}
\be\label{firstC}
\delta M=-T_{c} \delta S_{c}+\sum_i (\Omega^i_{c}-\Omega^i_{\infty})\delta J^i{\, +\, V_{c}\delta P}\,.
\ee
The minus sign in the first term on the right hand side arises because the surface gravity $\kappa_c$ of the de Sitter horizon is negative, while the corresponding temperature $T_c$ is taken to be {\em positive}.  The thermodynamic volume $V_c$ for the cosmological horizon is given by
\be\label{vdefC}
V_{c}= \int _{\infty} da r_e n_f ( \omega ^{ef}{-\omega_{dS}^{ef})} -\int _{dS} da r_e n_f \omega ^{ef} \,.
\ee
Finally, the Smarr relation that follows from (\ref{firstC}) by scaling is given by
\be\label{smarrC}
\frac{D-3}{D-2}M=-T_{c}S_{c}+\sum_i (\Omega^i_{c}-\Omega^i_{\infty})J^i{-\frac{2}{D-2}PV_{c}\,,}
\ee
We see from the first law \eqref{firstC} that the thermodynamic volume $V_c$ gives the variation in the mass assuming that the area of the cosmological horizon, as well as the angular momentum, is held fixed, which is distinct from the conditions defining the thermodynamic volume $V_h$.

The first law (\ref{first2})  discussed in the previous section, which concerns only the region between the black hole and cosmological horizons,
can now be understood as a consequence of subtracting  the formulas 
\eqref{firstBH} and \eqref{firstC}. 
We see from the explicit expressions of the thermodynamic volumes $V$ in (\ref{vdef}) and $V_h,V_c$ in (\ref{vdefBH}) and (\ref{vdefC}) that they satisfy the relation
\be\label{between}
V=V_{c}-V_{h}\,.
\ee
The Smarr formula (\ref{smarr2}) for the region between the horizons is similarly given by the difference between equations (\ref{smarrBH}) and (\ref{smarrC}).

\section{Thermodynamics of Kerr-deSitter black holes.}\label{kerrds}
We now turn to the example of Kerr-deSitter black holes.   The $D$-dimensional Kerr-(A)dS spacetimes \cite{GibbonsEtal:2004, GibbonsEtal:2005}  are solutions to the 
Einstein equations 
$$
R_{ab} =\frac{2\Lambda}{(D-2)}  g_{ab}
$$
that generalize the asymptotically-flat rotating black hole solutions  of \cite{MyersPerry:1986}.  The 
 thermodynamics for the Kerr-AdS case was studied in \cite{Gibbons:2005b}.
In the asymptotically dS case the metric 
in  `generalized' Boyer-Lindquist 
coordinates takes the form
\begin{eqnarray}\n{MPC}
ds^2&=&-W(1-g^2r^2)d t^2+\frac{2m}{U} \Bigl(Wd t-
\sum_{i=1}^{N}\frac{a_i\mu_i^2d \varphi_i}{\Xi_i}\Bigr)^{2} 
\nonumber\\
&+&\sum_{i=1}^{N}\frac{r^2+a_i^2}{\Xi_i}\,(\mu_i^2d \varphi_i^2
+d \mu_i^2)+\frac{Ud r^2}{X-2m}+\epsilon r^2 d \nu^2\nonumber\\
&+&
\frac{g^2}{W(1-g^2r^2)}\,
\Bigl(\sum_{i=1}^N\frac{r^2\!+\!a_i^2}{\Xi_i}\,\mu_i d \mu_i\!+\!
\epsilon r^2\nu d \nu\Bigr)^2,\ \ 
\end{eqnarray} 
where $2\Lambda = (D-1)(D-2)g^2$ and 
\begin{eqnarray}
W&=&\sum_{i=1}^N\frac{\mu_i^2}{\Xi_i}+\epsilon \nu^2\,,\quad 
X=r^{\epsilon-2}(1 -g^2 r^2)\prod_{i=1}^{N}(r^2+a_i^2)\,,\nonumber\\
U&=&\frac{Z}{1-g^2r^2}\,\Bigl(1-\sum_{i=1}^N\frac{a_i^2 \mu_i^2}{r^2+a_i^2}\Bigr)\,,\ \,\Xi_i=1+g^2a_i^2\,.\ \ 
\end{eqnarray}
Here $N\equiv [(D-1)/2]$, where $[A]$ means the integer part of $A$, and we 
have defined $\epsilon$ to be $1$ for $D$ even and $0$ for odd. The 
coordinates $\mu_i$ are not independent, but obey the constraint
\begin{equation}\label{constraint}
\sum_{i=1}^N\mu_{i}^2+\epsilon \nu^2=1\,.
\end{equation}

\subsection{Even dimensions}
In even dimensions ($D=2N+2$), the thermodynamic quantities are calculated as follows. 
For the cosmological horizon, we have   
\bea\label{TDevenH}
S_c&=& \ft14 \cA_{D-2}\, \prod_i \fft{r_c^2+a_i^2}{\Xi_i}=\frac{\cA_c}{4}\,,\nonumber\\ 
T_c&=& -\fft{r_c(1-g^2 r_c^2)}{2\pi} \sum_i \fft1{r_c^2+a_i^2} +
  \fft{1+g^2 r_c^2}{4\pi r_c}\,,\nonumber\\
\Omega_c^i &=& \fft{(1-g^2 r_c^2) a_i}{r_c^2 + a_i^2}\,,
\eea
while the `mass' and angular momenta read  
\be\label{Ji}
M=\frac{m \cA_{D-2}}{4\pi \prod_j \Xi_j}\sum_i \frac{1}{\Xi_i}\,,\quad 
J_i = \fft{m a_i \cA_{D-2}}{4\pi \Xi_i \prod_j\Xi_j}\,,
\ee 
where the cosmological horizon radius $r_c$ and black hole horizon radius $r_h$ are solutions to
\be\label{eq216}
2m = \frac{1}{r_c} (1-g^2 r_c^2)\prod_i (r_c^2 + a_i^2)=\frac{1}{r_h} (1-g^2 r_h^2)\prod_i (r_h^2 + a_i^2)\,.
\ee

The quantity 
${\cal A}_{D-2}$ is the volume of the unit-radius $(D-2)$-sphere, and is
given by
\be
{\cal A}_{D-2} = \fft{2\pi^{(D-1)/2}}{\Gamma[(D-1)/2]}\,.
\ee
Using the Smarr relation \eqref{smarrC} one finds that {the thermodynamic volume $V_c$ associated with the de Sitter horizon is given by}
\bea\label{VcKerr}
V_c &=& \frac{r_c \cA_c}{D-1}\left[1+\frac{1-g^2r_c^2}{(D-2)r_c^2}\sum_i \frac{a_i^2}{\Xi_i}\right]\\  
&=&{ \frac{r_c \cA_c}{D-1} +{8\pi\over (D-1)(D-2)}\sum_i a_iJ_i\,.  } \label{VcKerr2}
\eea
Given that the Smarr formula \eqref{smarrC} was derived from the first law \eqref{firstC}, it also follows that the  quantities \eqref{TDevenH}, \eqref{Ji}, and \eqref{VcKerr}
satisfy the cosmological horizon first law.  

Similarly, for the black hole horizon we have 
\bea\label{TDevenC}
S_h&=& \ft14 \cA_{D-2}\, \prod_i \fft{r_h^2+a_i^2}{\Xi_i}=\frac{\cA_h}{4}\,,\nonumber\\ 
T_h&=& \fft{r_h(1-g^2 r_h^2)}{2\pi} \sum_i \fft1{r_h^2+a_i^2} -
  \fft{1+g^2 r_h^2}{4\pi r_h}\,,\nonumber\\
\Omega_h^i &=& \fft{(1-g^2 r_h^2) a_i}{r_h^2 + a_i^2}\,,
\eea
while the thermodynamic volume calculated from \eqref{smarrBH} reads 
\bea\label{VBHKerr}
V_{h}& =& \frac{r_h \cA_h}{D-1}\left[1+\frac{1-g^2r_h^2}{(D-2)r_h^2}\sum_i \frac{a_i^2}{\Xi_i}\right]  \\\label{VBHKerr2}
&=&	{\frac{r_h \cA_h}{D-1}+{8\pi\over (D-1)(D-2)}\sum_i a_iJ_i\,.}
\eea
The horizon quantities \eqref{TDevenC}, \eqref{Ji}, and \eqref{VBHKerr} then 
satisfy the black hole horizon first law \eqref{firstBH}.  

The expressions for the thermodynamic volumes $V_c$  in  \eqref{VcKerr2} and $V_h$ in \eqref{VBHKerr2} have the same form as one another and also as the result for $V_h$ in the 
Kerr-AdS case \cite{CveticEtal:2010}.  As in  \cite{CveticEtal:2010} the first terms in these expressions for $V_c$ and $V_h$ are equal to geometric volumes, respectively denoted by 
\be
V_c^\prime=\frac{r_c\cA_c}{D-1}\,,\qquad V_h^\prime=\frac{r_h\cA_h}{D-1}\,,
\ee
contained within the horizon, which are obtained by integrating the full $D$-dimensional volume element over the region on a $t$ equals constant slice between $r=0$ and the horizon radius.\footnote{%
A similar formula holds also in odd dimensions, but the integration now proceeds between $r^2=-a^2_{min}$ where $a^2_{min}$ is the smallest among the values of the squares of the rotational parameters $a_i^2$. Note that in an even dimension such $a^2_{min}$ automatically equals zero.
}

We see from  \eqref{VcKerr2} and \eqref{VBHKerr2} that the thermodynamic volumes $V_c$ and $V_h$ differ from their geometric counterparts $V_c^\prime$ and $V_h^\prime$ by precisely the same amount.  It then follows that the thermodynamic volume $V=V_c-V_h$ that enters the first law \eqref{first2} and Smarr relation \eqref{smarr2} for the region between the two horizons is exactly equal to its geometric counterpart $V^\prime=V_c^\prime-V_h^\prime$, with
\be\label{VKerr}
V=V'=\frac{r_c \cA_c}{D-1}-\frac{r_h \cA_h}{D-1}\,.
\ee
In particular,
for the Schwarzschild-{dS spacetimes} we get the following `manifest' geometric relation
\be\label{VSchw}
V=\frac{\cA_{D-2}}{D-1}\left(r_c^{D-1}-r_h^{D-1}\right)\,.
\ee
{The} quantities \eqref{TDevenH}, \eqref{Ji}, \eqref{TDevenC}, and \eqref{VKerr}
satisfy the first law of black hole thermodynamics \eqref{first2} and the Smarr relation  \eqref{smarr2}.

\subsection{Odd dimensions}

In odd dimensions ($D=2N+1$), the following thermodynamic quantities get modified:  
the mass
\be
M=\frac{m \cA_{D-2}}{4\pi \prod_j \Xi_j}\left(\sum_i \frac{1}{\Xi_i}-\frac{1}{2}\right)\,,
\ee
the cosmological horizon entropy and temperature
\bea\label{TDodd}
S_c&=& \fft{\cA_{D-2}}{4r_c}\, \prod_i \fft{r_c^2+a_i^2}{\Xi_i}=\frac{\cA_c}{4}\,,\nonumber\\
T_c&=& -\fft{r_c(1-g^2 r_c^2)}{2\pi} \sum_i \fft1{r_c^2+a_i^2} +\fft1{2\pi r_c}\,,\nonumber
\eea
and the black hole horizon entropy and temperature
\bea\label{TDodd2}
S_h&=& \fft{\cA_{D-2}}{4r_h}\, \prod_i \fft{r_h^2+a_i^2}{\Xi_i}=\frac{\cA_h}{4}\,,\nonumber\\
T_h&=& \fft{r_h(1-g^2 r_h^2)}{2\pi} \sum_i \fft1{r_h^2+a_i^2} -\fft1{2\pi r_h}\,.
\eea
The other quantities, including $\Omega_h^i$ and $\Omega_c^i$, and $J_i$ remain of the same form, with  $m$ related to the cosmological and black hole horizon radii according to 
\be
2m = \fft1{r_c^2}\, (1-g^2 r_c^2)\prod_i (r_c^2 + a_i^2)
= \fft1{r_h^2}\, (1-g^2 r_h^2)\prod_i (r_h^2 + a_i^2)\,.
\ee
It is easy to verify that the thermodynamic volumes $V$, $V_c$ and $V_h$ again take the form \eqref{VKerr}, \eqref{VcKerr} and \eqref{VBHKerr} and that all the quantities satisfy the Smarr relations \eqref{smarr2}, \eqref{smarrBH}, \eqref{smarrC} and the first laws \eqref{first2}, \eqref{firstBH} and \eqref{firstC}.

\section{Isoperimetric inequalities}\label{isop}
\subsection{Euclidean space}
The isoperimetric inequality for
the volume ${\cal V}$ of a connected domain   
in Euclidean space ${\mathbb E}^{D-1}$ whose area
is ${\cal A}$ states that the ratio 
\ben\label{ratio}
R= \Bigl( \frac{(D-1){\cal V}\,}{{\cal A}_{D-2} } \Bigr )^{\fft1{D-1}}\,
  \Bigl(\fft{{\cal A}_{D-2}}{\cal A}\Bigr)^{\fft1{D-2}}
\een
obeys $R\leq 1$, with equality if and only if the domain is a standard round ball.  
That is, for a fixed volume, the area that surrounds the
volume is minimized when the volume is a ball.

\subsection{ Reverse isoperimetric inequalities for $V_h$ and $V_c$}

Kerr-AdS black holes have been shown to satisfy a `reverse'  isoperimetric inequality \cite{CveticEtal:2010}, with the thermodynamic volume bounded from below, rather than from above, in relation to the horizon area.   We find similar results in this subsection for the thermodynamic volumes $V_h$ and $V_c$ associated with the black hole and de Sitter horizons.   Both satisfy reverse isoperimetric inequalities in relation to the corresponding horizon areas.  Novel results will arise, however, when we consider the region between the horizons.  We will find that a true isoperimetric inequality bounds the thermodynamic volume $V$ of this region from above in terms of the horizon areas.

Consider first the black hole horizon where we take the volume $\cal V$ entering the isoperimetric inequality to be given by the thermodynamic volume of the black hole horizon
\be\label{VKerrISO2}
{\cal V}\equiv V_{h}=\frac{r_h \cA_h}{D-1}\left[1+\frac{1-g^2r_h^2}{(D-2)r_h^2}\sum_i \frac{a_i^2}{\Xi_i}\right]\,,
\ee
and the area $\cal A$ to be the black hole horizon area
\be\label{AKerrISO2}
{\cal A}=\cA_{h}=\left\{ \begin{array}{ll}
\frac{\cA_{D-2}}{r_h}\prod_i \frac{r_h^2+a_i^2}{\Xi_i}\,,& D\  \mbox{odd} \\
\cA_{D-2}\prod_i \frac{r_h^2+a_i^2}{\Xi_i}\,,&   D\  \mbox{even}\,.
\end{array}
\right.
\ee
The statement now is that the corresponding ratio $R$ defined in \eqref{ratio}, similar to the AdS case, satisfies the {\em reverse isoperimetric inequality}, 
\be\label{reverse}
R\geq 1\,,
\ee
with equality if and only if there is no rotation. This may be rephrased as stating that {\em for Kerr-dS spacetimes with a fixed thermodynamic volume $V_{h}$ 
the black hole entropy $\cA_h/4$ is maximized for Schwarzchild-dS.}  
As in the AdS case \cite{CveticEtal:2010},  we conjecture that a similar statement holds for any asymptotically dS black hole.

The proof of \eqref{reverse} for Kerr-dS follows closely the one for Kerr-AdS in \cite{CveticEtal:2010}. We define 
a dimensionless quantity
\be
z= \fft{(1-g^2 r_h^2)}{r_h^2}\,\sum_i\fft{a_i^2}{\Xi_i}\,,\label{zdef}
\ee
and consider $R^{D-1}$. In odd dimensions we find
\bea
R^{D-1} &=&r_h\, \Big[1+ \fft{z}{D-2}\Big]\, \Big[\fft1{r_h}\, \prod_i 
   \fft{(r_h^2+a_i^2)}{\Xi_i}\Big]^{-\fft1{D-2}}\,\nn\\
&=&  \Big[1+ \fft{z}{D-2}\Big]\, \Big[\prod_i
   \fft{(r_h^2+a_i^2)}{r_h^2\, \Xi_i}\Big]^{-\fft1{D-2}}\,\nn\\
&\ge& \Big[1\!+\! \fft{z}{D\!-\!2}\Big]\, \Big[\fft{2}{D\!-\!1}\Big(
\sum_i\fft1{\Xi_i} \!+\! \sum_i \fft{a_i^2}{r_h^2\, \Xi_i}\Big)
       \Big]^{-\fft{(D-1)}{2(D-2)}} \,\nn\\
&=& \Big[1+ \fft{z}{D-2}\Big]\, \Big[1+ \fft{2z}{D-1}
   \Big]^{-\fft{(D-1)}{2(D-2)}}\equiv F(z)\,,
\eea
where the inequality follows from the AG inequality (inequality relating the arithmetic and
geometric means)
\be
\bigl(\prod_i x_i\bigr)^{1/N} \le \frac{1}{N}\sum_i x_i
\ee
for non-negative quantities $x_i$, and the equality follows from \eqref{zdef}
 and the definition $\Xi_i=1+g^2a_i^2$.  
Noting that $F(0)=1$, and that
\be
\fft{d\log F(z)}{dz}= \fft{(D-3)\, z}{(D-2)(D-2+z)(D-1+2z)}\,,
\ee
which is positive for non-negative $z$ in $D>3$ dimensions, it follows
that $F(z)\ge1$, and hence the reverse isoperimetric inequality \eqref{reverse}
is satisfied by all odd-dimensional Kerr-dS black holes.

In even dimensions the calculation is similar.  One finds that
\bea
R^{D-1} &=&r_h\, \Big[1+ \fft{z}{D-2}\Big]\, \Big[\prod_i
   \fft{(r_h^2+a_i^2)}{\Xi_i}\Big]^{-\fft1{D-2}}\,\nn\\
&=&  \Big[1+ \fft{z}{D-2}\Big]\, \Big[\prod_i
   \fft{(r_h^2+a_i^2)}{r_h^2\, \Xi_i}\Big]^{-\fft1{D-2}}\,\nn\\
&\ge& \Big[1+ \fft{z}{D-2}\Big]\, \Big[\fft{2}{D-2}\Big(
\sum_i\fft1{\Xi_i} + \sum_i \fft{a_i^2}{r_h^2\, \Xi_i}\Big)
       \Big]^{-\ft12} \,\nn\\
&=& \Big[1+ \fft{z}{D-2}\Big]\, \Big[1+ \fft{2z}{D-2}
   \Big]^{-\ft12}\equiv G(z)\,.
\eea
Thus $G(0)=1$ and $d\log G(z)/dz\ge0$, and so again we conclude that
$R\ge1$.  Thus the reverse isoperimetric inequality holds for even-dimensional
Kerr-dS black holes also.
It further follows via direct substitution that identical results hold for the cosmological thermodynamic volume $V_c$ and horizon area $\cA_c$. 
One can therefore  state that {\em for Kerr-dS spacetimes having a fixed value of the  cosmological thermodynamic volume $V_{c}$, 
the cosmological horizon entropy $\cA_c/4$ is maximized by the Schwarzchild-dS spacetime.}

\subsection{ True isoperimetric inequality for the thermodynamic volume $V$ between the horizons}\label{inbetween}
Let us now focus on the thermodynamic volume $V$ of the region between the black hole and de Sitter horizon.
Recall that in section \ref{kerrds} we found that for Kerr-deSitter spacetimes $V$ coincides with the geometric volume $V^\prime$ between the horizons.
We now want to ask whether $V$ satisfies some sort of inequality with respect to the horizon areas.

Consider the Kerr-dS black hole in any dimension. 
For the purposes of establishing a bound, we will choose to work with the volume parameter $\cal V$ given by the geometric volume of the de Sitter horizon,
\be
\label{VKerrISO}
{\cal V}\equiv\frac{r_c \cA_c}{D-1}\geq V=\frac{r_c \cA_c}{D-1}-\frac{r_h \cA_h}{D-1}\,,
\ee
 which is manifestly greater than or equal to the thermodynamic volume $V$.
For the area $\cal A$ we take the area of the de Sitter horizon
\be\label{VKerrISO2}
{\cal A}\equiv \cA_c\leq \cA_c+\cA_h\equiv A
\ee
 which is less than or equal to the total area $A$ of the black hole and de Sitter horizons.
With these choices for  $\cal V$ and $\cal A$ we then find, following \cite{CveticEtal:2010}, that the ratio $R$ defined by \eqref{ratio} is given in all dimensions $D$ by
\be
R= \Big(\prod_i R_i\Big)^{-\fft1{(D-1)(D-2)}}\,,\quad R_i=\frac{1+a_i^2/r_c^2}{1+g^2a_i^2}\,.
\ee
Since one always has $\frac{1}{r_c^2}\geq g^2$ each term $R_i \geq 1$; consequently
\be\label{ISO}
R \leq 1\, .
\ee
On the one hand, this inequality might look trivial, since the volume and area
being compared are as in Euclidean space. The interesting thing is that this simple geometric
volume is what arises in the Smarr relation, and that this is true even with rotation.

The quantities $\cal V$ and $\cal A$, and hence also the quantities $V$ and $A$, therefore satisfy a {\em true isoperimetric inequality.  Equality , $R=1$, is attained in \eqref{ISO}} if and only there is no black hole (${\cal V}=V$), in which case $r_c^2=1/g^2$.
Since in this case we also have {${\cal A}=A$}   we may formulate the following result:
{\em For a fixed thermodynamic volume $V$ in between the black hole and dS horizons the total entropy $S=A/4$ is minimized if there is no black hole.} 
 Note that if instead choosing $({\cal V},{\cal A})$ as in (\ref{VKerrISO})  and (\ref{VKerrISO2}) we chose ${\cal V}=V$ and ${\cal A}=A$ the departure of $R$ from equality would be even more severe. 

We have seen that the thermodynamic volume $V$
between the two horizons  is the same as the
geometrical volume $V'$, even with rotation. So fixing $V$ is like fixing the
size of the observable universe. Hence
another way of stating the isoperimetric inequality  is that for 
a given size of the observable universe, the
entropy interior to the cosmological horizon is increased by adding a black hole,
even though the black hole pulls in the cosmological horizon \cite{Gibbons:1977mu}.
 By assuming that 
non-black hole inhomogeneities follow the same entropy rule,
this entropy increase has been used to estimate the probability that inflation occurs  
for a universe in ``the landscape" \cite{Albrecht:2004ke}. It is interesting that  $V'$ arises naturally
as part of the free energy balance between the cosmological and black hole horizons,
and that it applies to black holes with angular momentum. 

Lastly, we emphasize that this situation is opposite to asymptotically AdS or flat case \cite{CveticEtal:2010}, where the {\em reverse isoperimetric inequality} for the thermodynamic volume 
was proved for Kerr-AdS black holes in any dimension. We conjecture that this feature may remain valid for any asymptotically de Sitter black hole.

\section{Charged black holes}\label{charged}

{In this section, we will extend our analysis to various charged and rotating de Sitter black hole solutions.   It is straightforward to extend the Hamiltonian perturbation theory analysis of section~\ref{firstlaw} to include $U(1)$ charges in the first laws and Smarr relations holding for different sets of boundaries.  We then consider the examples of the Kerr-Newman-deSitter solution in $D=4$ and the de Sitter versions of the Chong-Cvetic-Lu-Pope charged, rotating solutions in minimal $D=5$ supergravity \cite{ChongEtal:2005b}, which includes a Chern-Simons interaction for the $U(1)$ gauge field.}

{For Kerr-Newman black holes, we find results that match those of section~\ref{kerrds}.  The black hole and cosmological thermodynamic volumes $V_h$ and $V_c$ each satisfy reverse isoperimetric inequalities.  The thermodynamic volume $V$ is again equal to the geometric volume $V^\prime$ for the region between the horizons and again satisfies a true isoperimetric inequality.  For Chong-Cvetic-Lu-Pope-dS black holes our results are less complete.  {Whereas we are able to show analytically that a similar set of inequalities hold subject to a certain limitation on the range of the charge and rotational parameters, in the general case we have only  numerical support for this claim.} Equality of the thermodynamic volume $V$ with the geometric volume $V^\prime$ between the horizons, however, is shown to hold over the entire parameter range.}

We find that the modifications to the first laws and Smarr relations for the various sets of boundaries are as one would expect.
For the region between the black hole and de Sitter horizon the new formulas are given by
\ba
0&=&T_{h} \delta S_{h}+T_{c} \delta S_{c}+\sum_j(\Phi_{h}^j-\Phi_{c}^j)\delta Q^j\nonumber\\
&&\qquad \qquad \quad +\sum_{i} (\Omega^i_{h}-\Omega^i_{c})\delta J^i-V \delta P\,,\quad \label{chargedfisrt}\\
0&=&T_{h}S_{h}+T_{c} S_{c}+\frac{D-3}{D-2}\sum_j(\Phi_{h}^j-\Phi_{c}^j)Q^j\,\nonumber\\
&&\qquad \qquad +\sum_{i}(\Omega_{h}^i-\Omega_{c}^i)J^i+\frac{2}{D-2}VP\,,\quad \label{chargedSmarr}
\ea
where $\Phi_h^i$ and $\Phi_c^i$ are the potentials for the electric (and magnetic) $U(1)$ charges evaluated at the black hole and de Sitter horizons.

For the regions stretching respectively from the black hole and de Sitter horizons out to infinity we have the first laws and Smarr relations
\ba
\delta M&=&T_{h} \delta S_{h}+\sum_j(\Phi_{h}^j-\Phi_{\infty}^j)\delta Q^j\nonumber\\
&&\qquad   +\sum_{i} (\Omega^i_{h}-\Omega^i_{\infty})\delta J^i+V_{h} \delta P\,,\quad \label{chargedfisrtBH}\\
\frac{D-3}{D-2}M&=&T_{h}S_{h}+\frac{D-3}{D-2}\sum_j(\Phi_{h}^j-\Phi_{\infty}^j)Q^j\,\nonumber\\
&&\quad +\sum_{i}(\Omega_{h}^i-\Omega_{\infty}^i)J^i-\frac{2}{D-2}V_{h}P\,,\quad \label{chargedSmarrBH}
\ea
and 
\ba
\delta M&=&-T_{c} \delta S_{c}+\sum_j(\Phi_{c}^j-\Phi_{\infty}^j)\delta Q^j\nonumber\\
&&\qquad   +\sum_{i} (\Omega^i_{c}-\Omega^i_{\infty})\delta J^i+V_{c} \delta P\,,\quad \label{chargedfisrtC}\\
\frac{D-3}{D-2}M&=&-T_{c}S_{c}+\frac{D-3}{D-2}\sum_j(\Phi_{c}^j-\Phi_{\infty}^j)Q^j\,\nonumber\\
&& +\sum_{i}(\Omega_{c}^i-\Omega_{\infty}^i)J^i-\frac{2}{D-2}V_{c}P\,.\quad \label{chargedSmarrC}
\ea
where additionally the quantities $\Phi^i_\infty$ are the values of the electric and magnetic potentials at infinity. As before,
subtracting respectively (\ref{chargedSmarrC}) from (\ref{chargedSmarrBH}) yields (\ref{chargedSmarr}), and 
subtracting respectively (\ref{chargedfisrtC}) from (\ref{chargedfisrtBH}) yields (\ref{chargedfisrt}).

\subsection{Kerr-Newman-dS black hole}
The $D=4$ Kerr-Newman-deSitter metric for a rotating, charged black hole with positive cosmological constant reads
\ba
ds^2&=&-\frac{\Delta}{\rho^2}\left(dt-\frac{a\sin^2\!\theta}{\Xi} d\varphi\right)^2+\frac{\rho^2}{\Delta} dr^2\nonumber\\
&&+\frac{\rho^2}{S} d\theta^2+\frac{S\sin^2\!\theta}{\rho^2}\left(adt-\frac{r^2+a^2}{\Xi}d\varphi\right)^2,\quad
\ea
where the various functions entering the metric are given by
\ba
\Delta&=&(r^2+a^2)(1-r^2g^2)-2mr+{\hat z}^2\\
 S&=&1+a^2g^2\cos^2\!\theta\,,\quad {\hat z}^2=q_e^2+q_m^2\,,\\
\rho^2&=&r^2+a^2\cos^2\!\theta\,,\quad \Xi=1+a^2g^2\, ,
\ea
and the vector potential is 
\be
\phi=-\frac{q_e r}{\rho^2}\Bigl(dt\!-\!\frac{a\sin^2\!\theta}{\Xi}d\varphi\Bigr)
-\frac{q_m\cos\theta}{\rho^2}\Bigl(adt\!-\!\frac{r^2\!+\!a^2}{\Xi}d\varphi\Bigr)\,.
\ee
The thermodynamic quantities in Kerr-Newman-AdS spacetimes were computed in \cite{CaldarelliEtal:2000}.  From those results we can infer the following formulas in the de Sitter case. For the cosmological horizon we find
\ba
S_{c}&=&\frac{\pi(r_{c}^2+a^2)}{\Xi}\,,\quad \Omega_{c}=\frac{a\Xi}{r_{c}^2+a^2}\\
T_{c}&=&-\frac{r_{c}\left(1-a^2g^2-3g^2r_{c}^2-\frac{a^2+{\hat z}^2}{r_{c}^2}\right)}{4\pi(r_{c}^2+a^2)}\,,\\
\Phi^{(e)}_{c}&=&\frac{q_e r_{c}}{r_{c}^2+a^2}\,,\quad 
\Phi^{(m)}_{c}=\frac{q_m r_{c}}{r_{c}^2+a^2}\,,
\ea
where for magnetic charge we take the magnetic potential analogous to the electric one  \cite{BarnichGomberoff:2008}.
{The mass, the electric and magnetic charges are given by
\be
M=\frac{m}{\Xi^2}\,,\quad Q_e=\frac{q_e}{\Xi}\,,\quad Q_m=\frac{q_m}{\Xi}\,,
\ee
while the angular momentum and the angular velocity at infinity read
\be
J=\frac{am}{\Xi^2}\,,\quad \Omega_\infty=ag^2\,.
\ee
From the Smarr formula \eqref{chargedSmarrC} we compute the cosmological thermodynamic volume 
\be
V_c=\frac{r_c \cA_c}{3}\left[1+\frac{a^2}{2\Xi r_c^2}\Bigl(1-g^2r_c^2+\frac{{\hat z}^2}{r_c^2+a^2}\Bigr)\right]\,,
\ee
and easily verify that the first law \eqref{chargedfisrtC} is satisfied.  Since $V_c$ in the charged case is greater than or equal to the uncharged one, while the $\cA_c$ has the same form, the reverse isoperimetric inequality for the cosmological thermodynamic volume \eqref{reverse} remains valid.} 

Similarly, for the black hole horizon we have the following expressions for the thermodynamic quantities
\ba
S_{h}&=&\frac{\pi(r_{h}^2+a^2)}{\Xi}\,,\quad \Omega_{h}=\frac{a\Xi}{r_{h}^2+a^2}\\
T_{h}&=&\frac{r_{h}\left(1-a^2g^2-3g^2r_{h}^2-\frac{a^2+{\hat z}^2}{r_{h}^2}\right)}{4\pi(r_{h}^2+a^2)}\,,\\
\Phi^{(e)}_{h}&=&\frac{q_e r_{h}}{r_{h}^2+a^2}\,,\quad 
\Phi^{(m)}_{h}=\frac{q_m r_{h}}{r_{h}^2+a^2}\,.
\ea
The black hole thermodynamic volume is then given by the expression
\be
V_{h}=\frac{r_h \cA_h}{3}\left[1+\frac{a^2}{2\Xi r_h^2}\Bigl(1-g^2r_h^2+\frac{{\hat z}^2}{r_h^2+a^2}\Bigr)\right]\,
\ee
which also satisfies the reverse isoperimetric inequality \eqref{reverse}.

Finally, the thermodynamic volume $V$ in between the horizons calculated from  \eqref{chargedSmarr} again takes a simple geometric form,
\be
V=\frac{1}{3}\left(r_{c} \A_{c}-r_{h}\A_{h}\right)\,. 
\ee
Defining the volume $\cal V$ and area $\cal A$ as in Section~\ref{inbetween} to be
\be
{\cal V}\equiv\frac{r_c \cA_c}{3}\geq V\,,\quad {\cal A}\equiv \cA_c\leq \cA_c+\cA_h\equiv A\,,
\ee
we find that the isoperimetric inequality \eqref{ISO} holds also for Kerr--Newman-deSitter black holes.

\subsection{Charged-dS rotating black hole in $D=5$}\label{d5gaugesec}
The rotating charged-AdS black hole with electromagnetic Chern-Simons term was constructed by 
Chong, Cvetic, Lu, and Pope \cite{ChongEtal:2005b}. {We find that a corresponding rotating charged-dS black hole exists as
an exact solution to the Einstein-Chern-Simons equations with positive cosmological constant, obtained via} analytic continuation $g\to ig$. This black hole solution reads  
\ba\label{Cvetic}
ds^2&=&-\frac{S\left[(1+g^2r^2)\rho^2 dt +2q\nu \right]dt}{\Xi_a \Xi_b \rho^2}+\frac{2q\nu\omega}{\rho^2}\nonumber\\
&&+\frac{f}{\rho^4}\left(\frac{Sdt}{\Xi_a \Xi_b}-\omega\right)^2+\frac{\rho^2 dr^2}{\Delta}
+\frac{\rho^2 d\theta^2}{S}\nonumber\\
&&+\frac{r^2+a^2}{\Xi_a}\sin^2\!\theta d\varphi^2+\frac{r^2+b^2}{\Xi_b}\cos^2\!\theta d\psi^2\,,\nonumber\\
\phi&=&\frac{\sqrt{3}q}{\rho^2}\left(\frac{Sdt}{\Xi_a\Xi_b}-\omega\right)\,,
\ea
where 
\ba\label{Delta}
\nu&=&b\sin^2\!\theta d\varphi+a\cos^2\!\theta d\psi\,,\nonumber\\
\omega&=&a\sin^2\!\theta \frac{d\varphi}{\Xi_a}+b\cos^2\!\theta \frac{d\psi}{\Xi_b}\,,\nonumber\\
S&=&1+a^2g^2\cos^2\!\theta+b^2g^2\sin^2\!\theta\,,\nonumber\\
\Delta&=&\frac{(r^2+a^2)(r^2+b^2)(1-g^2r^2)+q^2+2abq}{r^2}-2m\,,\nonumber\\
\rho^2&=&r^2+a^2\cos^2\!\theta +b^2\sin^2\!\theta\,,\nonumber\\
\Xi_a&=&1+a^2g^2\,,\quad \Xi_b=1+b^2g^2\,,\nonumber\\
f&=&2m\rho^2-q^2-2abqg^2\rho^2\,.   
\ea
{
In the case of a negative cosmological constant \cite{ChongEtal:2005b}, the solution describes an asymptotically AdS charged rotating black hole in a certain range of parameters $m, a, b$ and $q$. It is easy to check that there exists a range for these parameters in which the function $\Delta$ in Eq.~\eqref{Delta}, admits three positive real roots. Hence in this range the solution describes an asymptotically dS charged rotating black hole. To our knowledge this is the first demonstration of the existence of such black hole solutions.}

The thermodynamic quantities are \cite{ChongEtal:2005b}
\ba
M&=&\frac{\pi m(2\Xi_a+2\Xi_b-\Xi_a\Xi_b)-2\pi q ab g^2(\Xi_a+\Xi_b)}{4\Xi_a^2 \Xi_b^2}\,, \nonumber\\
J^a &=& \fft{\pi(2am+qb(1-a^2g^2)]}{4\Xi_a^2 \Xi_b}\,,\nonumber\\
J^b &=& \fft{\pi(2bm+qa(1-b^2g^2)]}{4\Xi_b^2 \Xi_a}\,,\nonumber\\
Q&=& \fft{\sqrt3\, \pi q}{4\Xi_a\Xi_b}\,.
\ea
For the black hole horizon we have 
\ba
T_h&=&\fft{r_h^4[1-g^2(2r_h^2+a^2+b^2)] -(ab+q)^2}{
  2\pi r_h[(r_h^2+a^2)(r_h^2+b^2)+abq]}\,,\nonumber\\ 
S_h&=&\fft{\pi^2[(r_h^2+a^2)(r_h^2+b^2)+a b q]}{
         2\Xi_a \Xi_b\, r_h}\,,\nonumber\\ 
\Phi_h&=& \fft{\sqrt3 \, q r_h^2}{(r_h^2+a^2)(r_h^2+b^2)+abq}\,,\nonumber\\
\Omega^a_h &=& \fft{a(r_h^2+b^2)(1-g^2 r_h^2)+b q}{
   (r_h^2+a^2)(r_h^2+b^2)+abq}\,,\nonumber\\
\Omega^b_h &=& \fft{b(r_h^2+a^2)(1-g^2 r_h^2)+a q}{
   (r_h^2+a^2)(r_h^2+b^2)+abq}\,.
\ea
The Smarr formula \eqref{chargedSmarrBH} then leads to an expression for the black hole thermodynamic volume
\ba
V_{h}&=&\frac{r_h \cA_h}{4}\Bigl[1+\frac{1-g^2r_h^2}{3r_h^2}\left(\frac{a^2}{\Xi_a}+\frac{b^2}{\Xi_b}\right)\nonumber\\
&&+\frac{q^2d^2+abq(d^2+r_h^2-r_h^2a^2b^2g^4)}{3\Xi_a\Xi_b r_h^2
[(r_h^2+a^2)(r_h^2+b^2)+abq]}\Bigr]\,.\quad
\ea
where $d^2=a^2+b^2+2a^2b^2g^2$. 
When we switch off one of the rotations ({\it e.g.} by setting $a=0$) the reverse isoperimetric inequality \eqref{reverse} still holds. 
{However, because of the complexity of the expressions we have not been able to analytically establish or disprove this result in the general case and we leave the question for future study. We have, nevertheless, confirmed the plausibility of this conjecture by a numerical study. 
}  

Similarly, for the cosmological horizon we have 
\ba
T_c&=&-\fft{r_c^4[1-g^2(2r_c^2+a^2+b^2)] -(ab+q)^2}{
  2\pi r_c[(r_c^2+a^2)(r_c^2+b^2)+abq]}\,,\nonumber\\ 
S_c&=&\fft{\pi^2[(r_c^2+a^2)(r_c^2+b^2)+a b q]}{
         2\Xi_a \Xi_b\, r_c}\,,\nonumber\\ 
\Phi_c&=& \fft{\sqrt3 \, q r_c^2}{(r_c^2+a^2)(r_c^2+b^2)+abq}\,,\nonumber\\
\Omega^a_c &=& \fft{a(r_c^2+b^2)(1-g^2 r_c^2)+b q}{
   (r_c^2+a^2)(r_c^2+b^2)+abq}\,,\nonumber\\
\Omega^b_c &=& \fft{b(r_c^2+a^2)(1-g^2 r_c^2)+a q}{
   (r_c^2+a^2)(r_c^2+b^2)+abq}\,,
\ea
{which gives 
\ba
V_{c}&=&\frac{r_c \cA_c}{4}\Bigl[1+\frac{1-g^2r_c^2}{3r_c^2}\left(\frac{a^2}{\Xi_a}+\frac{b^2}{\Xi_b}\right)\nonumber\\
&&+\frac{q^2d^2+abq(d^2+r_c^2-r_c^2a^2b^2g^4)}{3\Xi_a\Xi_b r_c^2
[(r_h^2+a^2)(r_c^2+b^2)+abq]}\Bigr]\,.\quad
\ea
}

{Finally, using the Smarr relation \eqref{chargedSmarr}, we find that the thermodynamic volume in between the horizons is
\ba
V&=&\frac{1}{4}(r_c \cA_c-r_h \cA_h)\nonumber\\
&=&\frac{\pi^2}{2\Xi_a \Xi_b}\left[(r_c^2\!+\!a^2)(r_c^2\!+\!b^2)\!-\!(r_h^2\!+\!a^2)(r_h^2\!+\!b^2)\right],\qquad \ \
\ea
which again, coincides with the naive geometric one. For $abq\geq0$ and taking 
\ba
{\cal V}&=&\frac{\pi^2}{2\Xi_a \Xi_b}(r_c^2\!+\!a^2)(r_c^2\!+\!b^2)\geq V\,,\nonumber\\
{\cal A}&=&\cA_c\leq \cA_c+\cA_h\,,
\ea 
we realize that the isoperimetric inequality \eqref{ISO} holds. {Numerical investigations indicate that this inequality survives provided we took ${\cal V}=V$ and ${\cal A}=\cA_c$ for all admissible parameters $a, b, m$ and $q$ for which the metric \eqref{Cvetic} describes an asymptotically dS charged rotating black hole.}   }

\section{Compressibility and speed of sound for black hole horizons}\label{compr}

Given the notion of the thermodynamic volume of a black hole, 
we can proceed to explore further thermodynamic 
properties, such as the compressibility and speed of sound, that make use of the volume. Here we will find the effective compressibility 
and speed of sound of the black hole.
For purposes of illustration, we shall restrict
ourselves to the case $D=4$ in which there is only one angular momentum. 
 The adiabatic compressibility of the black hole horizon is defined as
\ben\label{Compress}
 \beta_{S_h}= -\frac 1 {V_h}\left(\frac{\partial V_h}{\partial P}\right)_{{S_h},J},
\een
  $ \beta_{S_h}$ was computed 
 for rotating black holes in Anti-de Sitter in \cite{Dolan:2011b}. The same
formula continues to hold in de Sitter space-time
\ben
\beta_{S_h}=
\frac {36\,S_h j^4}
{\left( 3+{8\,p} \right)  \left(3+ 8\,p + 3\,j^2 \right)
\left( 6+16\,p +3\,j^2 \right)}\;,
\een 
where $p=P{S_h}$ and $j= \frac {2\pi J} {S_h}$, so in de Sitter $p$ is negative. 
One can check that at fixed entropy, the angular momentum is a maximum for the extremal case and that the compressibility is greatest when $T_h=0$. One finds that the angular momentum
is given by
\ben j^2_{max}=\Bigl(1+\frac {8p} 3\Bigr)(1+8p),
\een
the compressibility is
\ben
\left.\beta_S\right|_{extremal}=
\frac {2\,S \left( 1+8\,p \right) ^{2}}
{\left( 3+8\,p\right)^2  \left( 1+4\,p\right)}\;,
\een
and, for $p \le 0$, this is a maximum when $p=0$, which is
$\Lambda\rightarrow 0$.
Expressing the entropy in terms of the black hole mass $M$
the greatest possible compressibility is then
\ben \label{CompressExtreme}
\beta_{S_h}|_{p=0,j=1}=\frac {2 S_h}{9}= \frac{4\pi M^2 G^3}{9\, c^8},
\een
where  equation (\ref{Ji}) has been used with $D=4$ and Newton's constant and the speed of light
have been made explicit. For a black hole of a few solar masses, (\ref{CompressExtreme}) is three orders of magnitude
less than the compressibility due to neutron degeneracy pressure in
a neutron star of the same mass \cite{Dolan:2011b}: in other words
the black hole equation of state is very stiff.

A thermodynamic  \lq\lq speed of sound'', $v_s$, can be defined by making use
of the thermodynamic volume to define a black hole density $\rho=\frac {M}{V_h}$. Then
\ben v_s^{-2} = \left.\frac  {\partial \rho}{\partial P}\right|_{S,J}
=1+
\frac {9\,j^4}
{ \left( 6 + 16\,p + 3\,j^2 \right)^2}\;,
\een
Thus $0.9\le v_s^2 \le 1$. A non-rotating de Sitter black hole
always has $v_s^2=1$, whereas $v_s ^2$ is smallest as $j\rightarrow 1$ and 
$p\rightarrow 0$.

Of course this \lq\lq speed of sound'' is not associated with any kind of surface wave on the event horizon; rather it is a measure of the susceptibility of 
the black hole to changing its mass and volume when the pressure is changed,
keeping the area constant.


\section{Summary}\label{sum}

{We have shown that it is possible to identify a negative thermodynamic pressure with a positive cosmological constant,
allowing us to make sense of the thermodynamics of asymptotically de Sitter black holes. Both
the first law of black hole thermodynamics and the corresponding  Smarr relation were demonstrated to hold, and we illustrated
our resulting formulae by applying them to various rotating (and charged) black holes, including Kerr-dS black holes 
in all dimensions. We found that in all studied examples the thermodynamic volume in between the horizons} coincided with the naive geometric one, being equal to the  
difference of products of cosmological and black hole horizon radii and horizon areas.
We also studied the thermodynamics of the cosmological horizon and the black hole horizon separately. This allowed us to define the 
corresponding cosmological and black hole thermodynamic volumes. In all the examples we considered {(except for the doubly-rotating Einstein-Chern-Simons-dS black hole for which we have only  numerical evidence),} we showed that the reverse isoperimetric inequality 
holds provided we take the thermodynamic volume to be that of either the black hole or the cosmological horizon; should
we take the volume to be the naive geometric volume in between the horizons then the isoperimetric inequality holds.  We conjecture that these relations will remain valid for any asymptotically de Sitter black holes.

{An interesting case to consider for future study is the production of black holes in inflation. Without the black holes, the cosmological constant decays from the fluctuations in the scalar field that go into cosmological perturbations, and
the cosmological horizon shrinks accordingly. If black holes are produced as well, then provided
$\delta A_{h} >0$   the first law tells us that either, or both, of $(\cA_c ,\Lambda)$
must decrease.   Perhaps there is an isoperimetric inequality
that would have an interpretation in terms of such a process.}

{\em Note added.} We notice some overlap of our work with recent paper \cite{BhattacharyaLahiri:2013}.

\vspace{0.2cm}

\section{Acknowledgments}
We would like to thank Mirjam Cvetic for helpful discussions, Gary W. Gibbons for useful comments and reading the manuscript, and the anonymous referee for suggesting we investigate the Nariai limit.
This work was supported in part by the Natural Sciences and Engineering Research Council of Canada (Mann) and  was supported in part by NSF grant PHY-0555304 (Kastor and 
Traschen). Kastor, Kubiznak, and Traschen thank the Benasque Center for Theoretical 
Physics, at which this work was started, for their hospitality. Traschen thanks the Perimeter
Institute, at which this work was mostly completed, for their hospitality.

\appendix

\section{A remark on the Nariai limit and the thermodynamic volume}\label{Nari}
For asymptotically de Sitter black holes there exists an interesting so called {\em Nariai limit}, see \cite{
Nariai, GinspargPerry:1983} and also \cite{MannRoss:1995, BoothMann:1999}, in which the cosmological and black hole horizons coincide (an analogue of the mass=charge extremal Reissner--Nordstr\"om black hole). In this limit, interestingly, the region in between the horizons does not shrink to zero. In fact when one `zooms' into it, it can be demonstrated that (when also the time coordinate is rescaled) the metric remains finite and well defined. It is then an interesting question (for which we are grateful to the anonymous referee) to see what happens to the thermodynamic volume in between the horizons in such a limit.   

For simplicity we consider only the uncharged, non-rotating, 
Schwarzschild-dS solution
\ba
ds^2&=&-f d\tau^2+\frac{dr^2}{f}+r^2(d\theta^2+\sin^2\!\theta d\varphi^2)\,,\nonumber\\
f&=&1-\frac{2m}{r}-\frac{1}{3}\Lambda r^2\,,
\ea
and consider its Nariai limit, $m\rightarrow \frac{1}{3\sqrt{\Lambda}}$
and $r_c\rightarrow r_h$.  Following Ginsparg and Perry \cite{GinspargPerry:1983}
we define new coordinates
\be
 \epsilon\cos\chi = \sqrt{\Lambda}(r-r_0),\quad \psi=\sqrt{\Lambda}\,\epsilon\,t ,
\ee
and consider the limit $\epsilon\to 0$, while we set 
\be
9 \,m^2\Lambda = 1-3\,\epsilon^2\,,\quad 
r_0=\frac{1}{\sqrt\Lambda}\left(1-\frac{\epsilon^2}{6}\right)\,.
\ee
It follows that  
\be\label{using} 
r_c=r_0+\frac{\epsilon}{\sqrt{\Lambda}}\,,\quad
r_h=r_0-\frac{\epsilon}{\sqrt{\Lambda}}\,,
\ee
and the line element becomes that of $dS_2\times S^2$ 
\be\label{Nariai}
ds^2=\frac{1}{\Lambda} \bigl(  
d\chi^2 - \sin^2\chi\, d\psi^2 + d\theta^2 + \sin^2\theta \,d \varphi^2
\bigr)\,,
\ee
with $0\le\chi\le\pi$ and $0\le\psi<2\pi$. 
The original black hole horizon is now located at $\chi=0$ whereas the cosmological horizon occurs 
at $\chi=\pi$. We remark that the Nariai meric \eqref{Nariai} possesses   
a finite four-volume 
\be
V_4=\frac{16\pi^2}{\Lambda^2}\,.
\ee

Let us now turn to the thermodynamic volume in between the horizons, which in the original coordinates
reads 
\be\label{Vorig}
V=\frac{1}{3}\bigl(r_c \cA_c - r_h \cA_h\bigr)
=\frac{4\pi}{3}\bigl( r_c^3 - r_h^3\bigr)\,.
\ee
We shall now argue that such a volume vanishes, despite the fact that the region does not shrink to zero. Indeed, using \eqref{using} one finds 
\be 
V=\frac{4\pi}{3}\bigl( r_c^3 - r_h^3\bigr)\approx \frac{8\pi\,\epsilon}{\Lambda^{3/2}}  \, , 
\ee
which clearly vanishes as $\epsilon\rightarrow 0$.
This is consistent with the fact that, as noted previously, the four-volume $V_4$ remains finite. 
Since $0\le\psi<2\pi$ requires $0\le t<\frac{2\pi}{\sqrt{\Lambda}\,\epsilon}
=t_{max}$,  in Schwarzschild coordinates this finite 4-volume comes from
\be
 V_{t_{\max}}=\left(\frac{8\pi\,\epsilon}{\Lambda^{3/2}}\right)
\left( \frac{2\pi}{\sqrt{\Lambda}\,\epsilon}\right)=
\frac{16\pi^2}{\Lambda^2}\, .
\ee
As $\epsilon\to 0$ we have $V\rightarrow 0$ and $t_{max}\rightarrow \infty$\,.
Note finally that, if instead of taking the Nariai limit of the original expression \eqref{Vorig}, we started with the metric \eqref{Nariai} and calculated the thermodynamic volume in between the horizons (situated at $\chi=0$ and $\chi=\pi$), using for example the Smarr relation, we would find that such a volume 
necessarily vanishes as both horizons have the same areas and temperatures. In this sense the calculation of the volume of Nariai spacetime commutes with the limit and in both instances we recover a
vanishing quantity.



\providecommand{\href}[2]{#2}\begingroup\raggedright\endgroup

\end{document}